\documentclass[aps,preprint,superscriptaddress,showpacs]{revtex4}
\usepackage{epsfig}
\usepackage{psfrag}
\usepackage{amsfonts}

\usepackage{graphicx}% Include figure files
\usepackage{dcolumn}% Align table columns on decimal point
\usepackage{bm}% bold math

\newcommand{\fft}[2]{{\frac{#1}{#2}}}
\newcommand{\ft}[2]{{\textstyle{\frac{#1}{#2}}}}

\begin{document}

%%%%% title page %%%%%

\preprint{MCTP-08-56}

\title{
The Partition Function of a Wilson Loop in a Strongly Coupled $\mathcal N=4$
Supersymmetric Yang-Mills Plasma with Fluctuations}

\author{De-fu Hou}
\email{hdf@iopp.ccnu.edu.cn} \affiliation{Institute of Particle
Physics, Huazhong Normal University, Wuhan, 430079, China, Key
Laboratory of Quark and Lepton Physics (Huazhong Normal University),
Ministry of Education, China }

\author{James T. Liu}
\email{jimliu@umich.edu}
\affiliation{Michigan Center for Theoretical Physics,
Randall Laboratory of Physics,
The University of Michigan,
Ann Arbor, MI 48109--1040, USA}

\author{Hai-cang Ren}
\email{ren@mail.rockefeller.edu}
\affiliation{Physics Department, The Rockefeller University,
1230 York Avenue,
New York, NY 10021--6399}
\affiliation{Institute of Particle Physics,
Huazhong Normal University,
Wuhan, 430079, China}

\begin{abstract}

We examine the one-loop partition function describing the fluctuations
of the superstring in a Schwarzschild-AdS$_5\times S^5$ background.  On the
bosonic side, we demonstrate the one-loop equivalence of the Nambu-Goto
action and the Polyakov action for a general worldsheet, while on the
fermionic side, we consider the reduction of the ten-dimensional
Green-Schwarz fermion action to a two-dimensional worldsheet action.
We derive the partition functions of the worldsheets corresponding to
both straight and parallel Wilson lines.  We discuss the cancellation of the
UV divergences of the functional determinants in the thermal AdS background.

\end{abstract}


\maketitle

%%%%%%%%%%%%%%%%%%%%%%%%%%%%%%%%%%%%%%%%

\section{Introduction}

AdS/CFT duality \cite{Maldacena:1997re,Gubser:1998bc,Witten:1998qj}
opens a new avenue towards a qualitative or even semi-quantitative
approach to exploring the strong interaction regime of some quantum
field theories.  Based on the isomorphism between the conformal group
in four dimensions and the isometry group of AdS$_5$, the most
familiar example of AdS/CFT is the duality between string theory
formulated on AdS$_5\times S^5$ and $\mathcal N=4$ supersymmetric Yang-Mills
theory (SYM) on the boundary.  Working in a Poincar\'e patch of
AdS$_5$ of radius $L$, we may take the metric
\begin{equation}
ds^2=\frac{L^2}{z^2}(-dt^2+d\vec x^2+dz^2)+L^2d\Omega_5^2,
\label{ads}
\end{equation}
with boundary at $z=0$.  The $\mathcal N=4$ supersymmetry is required to
ensure conformal invariance at the quantum level, and the isometry group
of $S^5$ corresponds to the global $SU(4)$ $R$-symmetry of the SYM theory.
In particular, a weakly coupled string theory at large AdS radius
is dual to $\mathcal N=4$ SYM at large number of colors, $N_c$, and
large 't~Hooft coupling
\begin{equation}
\lambda \equiv g_{\rm YM}^2N_c = \fft{L^4}{\alpha'^2}.
\label{eq:ldef}
\end{equation}

In spite of the differences between $\mathcal N=4$ SYM and QCD, AdS/CFT
has been successfully applied to the understanding of a wide spectrum of
phenomenology of the quark gluon plasma (QGP) created by heavy ion
collisions. A nonzero temperature $T$ of the quantum field theory on the
boundary may be implemented by introducing a Schwarzschild black hole
in the AdS bulk.  This renders the metric to be
\begin{equation}
ds^2=\frac{L^2}{z^2}\left[-fdt^2+d\vec x^2+\frac{dz^2}{f}\right]+L^2d\Omega_5^2,
\label{adsbh}
\end{equation}
where $f=1-z^4/z_h^4$ with $z_h=1/(\pi T)$.  Among the salient results in
this regard are the equation of state \cite{Witten:1998}, the shear
viscosity \cite{Policastro:2001yc} and the energy loss of quarks
\cite{Herzog:2006gh,Liu:2006ug} to the leading order in large $\lambda$,
extracted from stress tensor correlators or from Wilson loops. The best
agreement between the predictions of AdS/CFT and the observations are
achieved in the range
\begin{equation}
5.5<\lambda<6\pi,
\end{equation}
of the 't~Hooft coupling, providing evidence for a strongly coupled QGP.

While the above results were originally derived at infinite 't~Hooft
coupling, it is important to examine the finite coupling corrections in
order to assess the robustness of the leading order results.  One source
of such corrections comes from the $\alpha'$ expansion of the low energy
effective string theory.  In the present maximally supersymmetric case,
the first correction terms enter at the $\alpha'^3$ level, corresponding
to $\mathcal O(\lambda^{-3/2})$ relative to the leading order in the
$\mathcal N=4$ SYM theory.  This correction has been calculated for the
equation of state \cite{Gubser:1998nz,Pawelczyk:1998pb}, the shear viscosity
\cite{Buchel:2004di,Buchel:2008sh}, the jet quenching parameter
\cite{Armesto:2006zv} and for the drag force \cite{VazquezPoritz:2008nw}.
In case of the Wilson loops, however, there is another
source of corrections, namely the quantum fluctuations of the string world
sheet which contribute a term of $\mathcal O(\lambda^{-1/2})$. In the case
of jet quenching and the drag force, this formally dominates over the
$\mathcal O(\lambda^{-3/2})$ corrections of
\cite{Armesto:2006zv,VazquezPoritz:2008nw} computed
from the $\alpha'^3$ modified supergravity background.

The expectation value of a Wilson loop operator in QCD is an important
quantity that probes the confinement mechanism in the hadronic phase and
meson melting and energy loss in the plasma phase. In the case of
$\mathcal N=4$ SYM, the theory is non-confining, and we are thus only
concerned with the plasma phase.  Following the AdS/CFT dictionary,
the Wilson loop expectation value is dual to the path integral of a
first-quantized superstring worldsheet in the AdS$_5\times S^5$ bulk
spanned by the loop on the boundary \cite{Maldacena:1998im,Rey:1998bq}.
In particular
\begin{equation}
W[C]\equiv \langle Pe^{i\oint_C dx^\mu A_\mu}\rangle
={\rm const.}\int[dX][d\theta]e^{\frac{i}{2\pi\alpha'}S[X,\theta]},
\label{wilson}
\end{equation}
where $A_\mu$ is the gauge potential, $S[X,\theta]$ is the
superstring action with bosonic (fermionic) coordinates $X$'s
($\theta$'s) and the symbol $P$ indicates path ordering. The strong
coupling expansion of the SYM Wilson loop corresponds to the
semi-classical expansion of the string theory action, where the
inverse string tension $\alpha'$ is related to the 't~Hooft coupling
$\lambda$ through (\ref{eq:ldef}).  We have, apart from an additive
constant,
\begin{equation}
\ln W[C]=i\sqrt{\lambda}\left(S[\bar X,0]
+\frac{S_1[C]}{\sqrt{\lambda}}+\cdots\right),
\end{equation}
where $\bar X$ denotes the solution of the classical equation of
motion, $S_1[C]$ is the one-loop correction of the worldsheet
fluctuations and $\ldots$ represents the multi-loop contributions.

The classical solutions for
simple geometries of the Wilson loop $C$ (a straight line, a circle, a pair
of parallel lines, {\it etc.}) have been obtained explicitly for both
vacuum AdS (\ref{ads}) and Schwarzschild-AdS (\ref{adsbh}) metrics. The
one-loop fluctuations of the superstring in the vacuum AdS$_5\times S^5$
background have been discussed extensively in \cite{Forste:1999qn} and
\cite{Drukker:2000ep}, and fluctuations in a Schwarzschild-AdS background
were considered in \cite{Greensite:1999jw} and \cite{Naik:1999bs}.  In
this work, we complete the analysis of the one-loop string partition function
in the Schwarzschild-AdS background using a systematic treatment of both
bosonic and fermionic worldsheet fluctuations along the lines of
\cite{Forste:1999qn,Drukker:2000ep}.  On the gauge theory side, these
results allow for a systematic treatment of the Wilson loop at a nonzero
temperature.

In order to properly treat the string worldsheet in a Ramond-Ramond background,
we use the Green-Schwarz action for $S[X,\theta]$ of (\ref{wilson}).  The
Green-Schwarz action was originally formulated in a ten-dimensional
Minkowski background in \cite{Green:1983wt,Green:1983sg} and subsequently
generalized to the full IIB supergravity background in \cite{Grisaru:1985fv}
using superspace techniques.  While formally elegant, this IIB superspace
formulated action is not so practical to work with.  Instead, a complementary
approach to constructing the full Green-Schwarz action in an
AdS$_5\times S^5$ background was taken in \cite{Metsaev:1998it} using
supercoset methods based on the symmetries of the background.  Even in
this background, the action is rather complicated, although simplifications
may be obtained after gauge fixing the $\kappa$-symmetry
\cite{Pesando:1998fv,Kallosh:1998nx,Kallosh:1998ji}.

Since we are interested in a Schwarzschild-AdS background, the construction
of \cite{Metsaev:1998it} is not directly applicable.  However, all that is
needed for obtaining the one-loop partition function is the form of the
Green-Schwarz action up to quadratic order in the fermions.  This was obtained
in component form in \cite{Cvetic:1999zs} by dimensionally reducing and
T-dualizing the eleven-dimensional supermembrane action.  At quadratic order,
the action is a straightforward generalization of
\cite{Green:1983wt,Green:1983sg}, using the pullback of the IIB
supercovariant derivative in place of the ordinary covariant derivative.
Furthermore, it is valid in any background satisfying the IIB equations of
motion.  Although the black hole background will break the supersymmetry,
the $\kappa$ symmetry still holds trivially to quadratic order (and is
expected to hold to all orders) in the fermions.

For the simple Wilson loops considered in this paper, a straight line and a
pair of parallel lines, the gauge fixed
ten-dimensional Green-Schwarz fermion action reduces to a sum
of worldsheet fermionic actions of an equal mass that depends only on the
worldsheet curvature. The masses of the bosonic fluctuations, on the other
hand, receive nontrivial contributions from the Weyl tensor of the
non-extremal metric (\ref{adsbh}). However they do not add any new UV
divergences. Upon computing the first quantum correction to the path
integral, the partition function consists of a set of $1\times 1$ functional
determinants which may be evaluated by the recently suggested method of
\cite{Kruczenski:2008zk}.

This paper is organized as follows. In the next section, we examine the
relation between the fluctuations of the Nambu-Goto action and the Polyakov
action for a general worldsheet, and show that they are equivalent at the
one-loop level.  We furthermore reduce the ten-dimensional Green-Schwarz
fermion action to a worldsheet fermion action.  Following this general
analysis, we turn to the computation of the Wilson loops.  The partition
function corresponding to a straight line Wilson loop is derived in
section III and that corresponding to the parallel lines Wilson loop
is derived in the section IV.  In the final section, we examine the
cancellation of the UV divergences of the functional determinants in
the non-extremal background as well as the computation algorithm of these
determinants and some open issues.  Additional technical details on
longitudinal bosonic fluctuations are given in Appendix A, and we give
our fermion conventions in Appendix B.

Our notation largely follows that of
\cite{Drukker:2000ep}. Without a special declaration, Latin letters with
hats, $\hat a, \hat b, \hat c,\ldots$, labels the 10 Lorentz components
of the target space metric, while Latin letters without hats,
$a, b, c,\ldots$, pertain to the AdS$_5$ sector only. The Greek letters,
$\mu, \nu, \rho, \lambda,\ldots$, label the 10 dimensional curved-space
coordinates.  On the string worldsheet, the Latin letters $i, j, k, l,\ldots$,
label the curved-space coordinates, while the Greek letters,
$\alpha, \beta,\ldots$, stand for Lorentz indexes.

%%%%%%%%%%%%%%%%%%%%%%%%%%%%%%%%%%%%%%%%

\section{General formulation}

A careful study of the one-loop partition function of the Green-Schwarz
superstring in AdS$_5\times S^5$ was done by Drukker, Gross and Tseytlin
in \cite{Drukker:2000ep}.  The basic procedure is to start with a classical
string configuration, and then to develop the quadratic fluctuations about
this classical worldsheet configuration.  At this order, the bosonic and
fermionic fluctuations decouple, and hence can be considered separately.

\subsection{Bosonic contribution}

The bosonic part of the string action describes the embedding
$X^\mu(\sigma^i)$ of a two-dimensional worldsheet (parameterized by
$\sigma^i$ with $i=0$, $1$) into $D$-dimensional curved spacetime endowed
with a metric
\begin{equation}
ds^2=G_{\mu\nu}dX^\mu dX^\nu.
\label{target}
\end{equation}
This embedding may be accomplished by either the Nambu-Goto or the
Polyakov action.  The Nambu-Goto action is given by
\begin{equation}
S_{\rm NG}[X^\mu]=\frac{1}{2\pi\alpha'}\int d^2\sigma\sqrt{|g|}
\label{NambuGoto}
\end{equation}
where $g$ is the determinant of the induced metric
\begin{equation}
g_{ij}=G_{\mu\nu}\frac{\partial X^\mu}{\partial\sigma^i}\frac{\partial X^\nu}{\partial\sigma^j}.
\label{eq:induced}
\end{equation}
In this case, the action principle directly corresponds to the extremization
of the worldsheet volume, and is a straightforward generalization of the
relativistic particle action.

The Nambu-Goto action is nonlinear and often
difficult to work with because of the presence of the square root.  For
this reason, the approach of the Polyakov action is often preferred.  This
action is given by
\begin{equation}
S_P[g_{ij},X^\mu]=\frac{1}{4\pi\alpha'}\int d^2\sigma\sqrt{|g|}g^{ij}G_{\mu\nu}
\frac{\partial X^\mu}{\partial\sigma^i}\frac{\partial X^\nu}{\partial\sigma^j},
\label{Polyakov}
\end{equation}
where this time the worldsheet metric $g_{ij}(\sigma^k)$ is taken to be
independent of $X^\mu$.  It is well known that the Nambu-Goto and Polyakov
actions are equivalent at the tree-level.  To see this, one notes that the
classical equation of motion for $g_{ij}$ obtained from $S_P$ gives a result
which is conformally equivalent to the induced metric (\ref{eq:induced}).
Substituting this back in then reproduces the Nambu-Goto action,
(\ref{NambuGoto}).

It is generally expected that the classical equivalence of (\ref{NambuGoto})
and (\ref{Polyakov}) will be maintained at the quantum level, although
actually proving this equivalence does not appear to be straightforward.
One-loop equivalence of the actions for a Minkowski target space was
demonstrated \cite{Fradkin:1982ge}.  Furthermore, both forms of the action
were considered in \cite{Drukker:2000ep} when computing the bosonic
fluctuations of the superstring in a curved target space.  For non-trivial
string configurations, the results (when written in terms of bosonic
determinants) do not appear manifestly identical, and hence equivalence of
the actions implies the existence of hidden relations among the determinants.

Here we revisit the quadratic bosonic fluctuations considered in
\cite{Drukker:2000ep} and in addition demonstrate the equivalence of
the Nambu-Goto and Polyakov actions at the one-loop order.  We begin
with the Nambu-Goto action (\ref{NambuGoto}) and first consider the
background worldsheetconfiguration.  This is given by $\bar
X^\mu(\sigma)$, which satisfies the classical equation of motion
\begin{equation}
g^{ij}\nabla_i\eta_j^\mu=0,
\label{EOM}
\end{equation}
where $\eta_i^\mu\equiv\partial\bar X^\mu/\partial\sigma^i$ is both
a target space vector and a worldsheet vector.  Its covariant
derivative takes the form
\begin{equation}
\nabla_i\eta_j^\mu=\frac{\partial \eta_j^\mu}{\partial\sigma^i}
+\Gamma_{\rho\lambda}^\mu\eta_i^\rho\eta_j^\lambda
-\gamma_{ij}^k\eta_k^\mu
\end{equation}
with $\Gamma_{\rho\lambda}^\mu$ the Christoffel connection of the
target space and $\gamma_{ij}^k$ that of the worldsheet . The
background worldsheet  metric reads
\begin{equation}
\bar g_{ij}=G_{\mu\nu}\eta_i^\mu\eta_j^\nu.
\end{equation}
and the classical value of the action is simply the area of the minimal
surface described by $\bar X^\mu(\sigma)$
\begin{equation}
S_0=\fft1{2\pi\alpha'}\int d^2\sigma\sqrt{|\bar g|}.
\end{equation}

We now consider the fluctuations $X^\mu=\bar X^\mu+\delta X^\mu$ about
the classical background.  In order to do so, we expand the Nambu-Goto
action, (\ref{NambuGoto}), to the quadratic order in $\delta X^\mu$.
\begin{equation}
\delta g_{ij}\equiv g_{ij}-\bar g_{ij}=\delta_1 g_{ij}+\delta_2 g_{ij}
\end{equation}
where $\delta_1g_{ij}$ and $\delta_2g_{ij}$ denote terms linear and
quadratic in $\delta X^\mu$, respectively
\begin{eqnarray}
\delta_1g_{ij}&=&2G_{\mu\nu}\eta_{(i}^\mu\nabla_{j)}^{\vphantom\mu}
\delta X^\nu,\nonumber\\
\delta_2g_{ij}&=&G_{\mu\nu}\nabla_i\delta X^\mu \nabla_j\delta X^\nu -
R_{\rho\mu\lambda\nu}\eta_i^\rho \eta_j^\lambda\delta X^\mu\delta X^\nu.
\label{deltag1g2}
\end{eqnarray}
Introducing the target space $D$-beins $G_{\mu\nu}=E_\mu^aE_\nu^b\eta_{ab}$
allows us to define target space tangent coordinates
$\zeta^a=E_\mu^a\delta X^\mu$.  In this case, the above fluctuations may
be re-expressed as
\begin{eqnarray}
\delta_1g_{ij}&=&2E_\mu^a\eta_{(i}^\mu\nabla_{j)}^{\vphantom\mu}
\zeta_a,\nonumber\\
\delta_2g_{ij}&=&\nabla_i\zeta^a \nabla_j\zeta_a -
R_{\rho\mu\lambda\nu}\eta_i^\rho\eta_j^\lambda E_a{}^\mu E_b{}^\nu
\zeta^a\zeta^b.
\end{eqnarray}
It follows that, to quadratic order in $\delta X^\mu$, the Nambu-Goto
action becomes
\begin{equation}
S_{\rm NG}=S_0+S_{\rm NG}^{(2)}
=S_0+\frac{1}{2\pi\alpha'}(I-J),
\label{eq:sng2}
\end{equation}
where
\begin{eqnarray}
I&=&\frac{1}{2}\int d^2\sigma\sqrt{|\bar g|}\bar g^{ij}\delta_2g_{ij},
\nonumber\\
J&=&\frac{1}{8}\int d^2\sigma\sqrt{|\bar g|} (\bar g^{ik}\bar g^{jl}
+\bar g^{il}\bar g^{jk}-\bar g^{ij}\bar g^{kl})\delta_1g_{ij}\delta_1g_{kl}.
\label{I1I2}
\end{eqnarray}
Note that, at this quadratic order, we may symbolically write
\begin{equation}
I = \frac{1}{2}\tilde\zeta{\cal A}\zeta,\qquad
J = \frac{1}{2}\tilde\zeta{\cal A}^\prime\zeta,
\end{equation}
where ${\cal A}$ and ${\cal A}^\prime$ are $D\times D$ matrices of
functional operators.

We now decompose the fluctuation $\delta X^\mu$ into longitudinal
and transverse components, $\delta_{\rm l}X^\mu$ and $\delta_{\rm tr}X^\mu$,
respectively, according to
\begin{equation}
\delta_{\rm l}X^\mu = \eta_k^\mu\epsilon^k,\qquad
G_{\mu\nu}\delta_{\rm tr}X^\mu\delta_{\rm l}X^\nu = 0.
\label{longitudinal}
\end{equation}
Here $\epsilon^k$ is a worldsheet  vector specifying the two
independent longitudinal fluctuations.  It follows from
(\ref{deltag1g2}) that a longitudinal fluctuation generates a world
sheet diffeomorphism
\begin{equation}
\delta_1g_{ij}=\nabla_i\epsilon_j+\nabla_j\epsilon_i,
\end{equation}
and hence is a zero mode of the fluctuation action (\ref{eq:sng2}) and
(\ref{I1I2}).  (See Appendix A for a detailed proof.)  In this case, we
may perform a local $SO(1,D-1)$ rotation ${\cal R}$ to highlight the
transverse fluctuations
\begin{equation}
\left( \matrix {\zeta_{\rm tr}\cr \zeta_{\rm l}}\right)={\cal R}\zeta,
\end{equation}
so that
\begin{equation}
I-J = \frac{1}{2}\tilde\zeta({\cal A}-{\cal A}^\prime)\zeta,
\end{equation}
where
\begin{equation}
{\cal A}-{\cal A}^\prime={\cal R}^{-1}\left(\matrix{A_{\rm
tr}&0\cr 0&0\cr}\right){\cal R}. \label{NG}
\label{mtr}
\end{equation}
with $A_{\rm tr}$ a $(D-2)\times(D-2)$ matrix of functional operators.
This highlights the well known feature that only the $D-2$ transverse
degrees of freedom are physical in the fundamental string action.

Having examined the Nambu-Goto form of the string action, we now
turn to the Polyakov action, (\ref{Polyakov}).  In this case, the
solution to the equations of motion gives rise to the same
worldsheet embedding profile $\bar X^\mu$ and, up to a conformal
factor, the same induced metric $\bar g_{ij}$.  In conformal gauge,
where the fluctuation of $g_{ij}$ is restricted to $\delta g_{ij}
\propto \bar g_{ij}$, the expansion of the Polyakov action to
quadratic order in $\delta X^\mu$ yields
\begin{equation}
S_{\rm P}=S_0+S_{\rm P}^{(2)}=S_0+\fft1{2\pi\alpha'}I,
\end{equation}
where $I$ was given previously in (\ref{I1I2}).  At this point, we note that
the quadratic fluctuations of the Nambu-Goto and Polyakov actions differ in
that $S_{\rm NG}$ contains the additional $J$ given in (\ref{I1I2}).  On
the other hand, gauge fixing the Polyakov action gives rise to a Fadeev-Popov
ghost determinant, which is not present in the above treatment of the
Nambu-Goto action.  In order to demonstrate the equivalence of these two
formulations, we must show that evaluation of this ghost determinant gives
the same result as the addition of the $J$ term in $S_{\rm NG}$.

Accounting for the ghosts, the partition function for bosonic fluctuations
in the Polyakov action is given by the path integral
\begin{equation}
Z_{\rm bs} = (\det A_{\rm gh})^{1/2}\int [d\zeta]e^{iS_P^{(2)}}
={\rm (const.)} \frac{(\det A_{\rm gh})^{1/2}}
{(\det{\cal A})^{1/2}},
\label{conformal}
\end{equation}
where the $2\times 2$ ghost operator is
\begin{equation}
A_{\rm gh}=-\nabla^j\nabla_j-\frac{1}{2}R^{(2)}.
\end{equation}
Here $\nabla_j$ is the covariant derivative acting on a worldsheet
Lorentz vector and $R^{(2)}$ is the worldsheet  scalar curvature.
Note that we have dropped the path integral over the conformal
factor, which only decouples in the critical dimension.

For simple embeddings, such as the minimal surface corresponding to a straight
Wilson line or that of a circular Wilson loop on the AdS$_5$ boundary,
explicit computations indicate that the bosonic determinant factorizes
according to
\begin{equation}
\det{\cal A}=\det{A_{\rm tr}}\det{A_{\rm gh}}.
\label{equiv}
\end{equation}
This, however, is not obvious with more complicated embeddings, such
as the worldsheet  generated by parallel lines on the boundary. In
the rest of this subsection, we give an explicit demonstration that
this is the case in general.  Moreover, we note that this
factorization does not require a simple relationship among the
eigenvalues of ${\cal A}$, $A_{\rm tr.}$ and $A_{\rm gh.}$.

To begin with, we note that $J$, given in (\ref{I1I2}), depends only on the
traceless part of $\delta_1g_{ij}$.  In other words,
\begin{equation}
J = \frac{1}{4}\int d^2\sigma\sqrt{|\bar g|}\bar g^{ik}\bar g^{jl}
h_{ij}h_{kl},
\end{equation}
where $h_{ij}$ is the traceless part of $\delta_1g_{ij}$, given by
\begin{equation}
h_{ij}\equiv \delta_1g_{ij}-\frac{1}{2}\bar g_{ij}\bar g^{kl}\delta_1g_{kl}.
\end{equation}
Decomposing $\delta_1g_{ij}$ into longitudinal and transverse components
\begin{equation}
\delta_1g_{ij}=\nabla_i\epsilon_j+\nabla_j\epsilon_i+\delta g_{ij}^{\rm tr},
\end{equation}
allows us to write
\begin{equation}
h_{ij}=\nabla_i\epsilon_j+\nabla_j\epsilon_i-\bar g_{ij}\bar g^{kl}
\nabla_k\epsilon_l+h_{ij}^{\rm tr},
\end{equation}
where $h_{ij}^{\rm tr}$ are both transverse and traceless.  While there
are $D-2$ transverse components $\zeta_{\rm tr}^a$, we note that there
are only two independent components of $h_{ij}^{\rm tr}$.  In this case,
we may find a worldsheet vector $\chi^k$ that satisfies
\begin{equation}
\nabla_i\chi_j+\nabla_j\chi_i-\bar g_{ij}\bar g^{kl}\nabla_k\chi_l
=h_{ij}^{\rm tr}.
\label{eq:dchi}
\end{equation}
The formal solution may be expressed as
\begin{equation}
\chi^i=B^i{}_a\zeta_{\rm tr}^a,
\end{equation}
where $B$ is a nonlocal linear operator representing the inversion of
(\ref{eq:dchi}). We thus end up with
\begin{equation}
J=\frac{1}{2}\int d^2\sigma\sqrt{|\bar g|}
\Big(\nabla_iu^j\nabla^iu_j-\frac{1}{2}R^{(2)}u^iu_i\Big)=\frac{1}{2}
\int d^2\sigma u^i A_{\rm gh.}u_i,
\end{equation}
where $u^i=\epsilon^i+\chi^i$.

Following the definition of (\ref{I1I2}), the above expression for $J$
implies that the matrix operator ${\cal A}^\prime$ has the form
\begin{equation}
{\cal A}^\prime={\cal R}^{-1}\left(\matrix{1&\tilde B\cr 0&1\cr}\right)
\left(\matrix{0&0\cr 0&A_{\rm gh.}\cr}\right)
\left(\matrix{1&0\cr B&1\cr}\right){\cal R}.
\label{NGp}
\end{equation}
By adding (\ref{NG}) and (\ref{NGp}), we may obtain an expression for
$\mathcal A$
\begin{equation}
{\cal A}={\cal R}^{-1}\left(\matrix{1&\tilde B\cr 0&1\cr}\right)
\left(\matrix{A_{\rm tr.}&0\cr 0&A_{\rm gh.}\cr}\right)
\left(\matrix{1&0\cr B&1\cr}\right){\cal R}.
\end{equation}
As a result, the factorization (\ref{equiv}) follows, and we are left with
the bosonic partition function
\begin{equation}
Z_{\rm bs}=(\mbox{const.})(\det{A_{\rm tr}})^{-1/2},
\label{eq:zbs}
\end{equation}
This result is independent of whether the Nambu-Goto or the Polyakov
action was taken as the starting point of the computation.  Finally,
we note that the nonlocal operator $B$ vanishes for the long string and
circular loop configurations.  This is the reason why the factorization
(\ref{equiv}) was already manifest for such simple embeddings.  However the
operator $B$ may be highly nontrivial for generic string configurations.
In such cases, we often find it more convenient to work with the Nambu-Goto
action, where we do not have to concern ourselves with the added complication
of gauge fixing and ghosts.

\subsection{Fermionic contribution}

For the bosonic fluctuations, we have shown above that the
equivalence of working with the Nambu-Goto and Polyakov actions
holds for arbitrary target spacetimes.  We note, however, that
because of the UV divergence, the path integral over the conformal
factor only decouples in the critical dimension.  This is not a
concern for us since our primary interest is to consider the
superstring in the critical dimension, and in particular in a Schwarzschild
AdS$_5\times S^5$ background.

To calculate the fermionic contribution to the one-loop partition function,
we work with the Green-Schwarz action up to quadratic order in the fermions.
Turning on only a background metric and 5-form field strength, the quadratic
action takes the form \cite{Cvetic:1999zs}
\begin{equation}
S_F=\fft{i}{2\pi\alpha'}\int d^2\sigma\sqrt{|\bar g|}
[\bar\theta^1P_-^{ij}\eta_i^\mu\Gamma_\mu(D_j\theta)^1
+\bar\theta^2P_+^{ij}\eta_i^\mu\Gamma_\mu(D_j\theta)^2],
\label{eq:gsact}
\end{equation}
where the fermionic coordinates $\theta^I$ ($I=1,2$) are 16 component
positive chirality Majorana-Weyl spinors.  We have also introduced the
worldsheet projection tensors
\begin{equation}
P_\pm^{ij}=\bar g^{ij}\pm\frac{\epsilon^{ij}}{\sqrt{|\bar g|}},
\label{projection}
\end{equation}
and the pullback of the IIB supercovariant derivative
\begin{equation}
D_i^{IJ}=\delta^{IJ}\mathcal D_i+\fft1{16\cdot5!}\epsilon^{IJ}
F_{\nu_1\cdots\nu_5}
\Gamma^{\nu_1\cdots\nu_5}\Gamma_\mu\eta_i^\mu,
\label{eq:scder}
\end{equation}
where
\begin{equation}
\mathcal D_i=\fft\partial{\partial\sigma^i}+\fft14\eta_i^\mu
\Omega_\mu^{\hat a\hat b}\Gamma_{\hat a\hat b}.
\end{equation}
Our notation follows that of \cite{Drukker:2000ep,Metsaev:1998it}, and is
detailed in Appendix B.

The $S^5$ reduction of the IIB theory is given by
\begin{equation}
ds_{10}^2=ds_5^2+L^2d\Omega_5^2,\qquad
F_5=\fft4L(1+*)d\Omega_5,
\end{equation}
and leads to a natural $5+5$ split of spacetime.  This motivates a split of
the ten-dimensional Dirac matrices into two sets of commuting $16\times 16$
Dirac matrices, $\gamma^a$ and $\gamma^{a'}$, satisfy the relations
\begin{equation}
\{\gamma^a,\gamma^b\}=2\eta^{ab},\qquad
\{\gamma^{a'},\gamma^{b'}\}=2\delta^{a'b'},
\label{anti12}
\end{equation}
where $a,b$ are local Lorentz indexes on AdS$_5$, and $a',b'$ that on $S^5$.
We now introduce worldsheet pullbacks of the Dirac matrices
\begin{eqnarray}
\rho_i&=&\eta_i^\mu (E_\mu^a\gamma_a+iE_\mu^{a'}\gamma_{a'}),\nonumber\\
\bar\rho_i&=&\eta_i^\mu (E_\mu^a\gamma_a-iE_\mu^{a'}\gamma_{a'}).
\end{eqnarray}
It follows from (\ref{anti12}) that
\begin{equation}
\bar\rho_i\rho_j+\bar\rho_j\rho_i=2\bar g_{ij},
\label{clifford}
\end{equation}
where, as we recall, $\bar g_{ij}$ is the background worldsheet metric.

Making this $5+5$ split explicit, we find that the action (\ref{eq:gsact})
takes the form
\begin{equation}
S_{\rm F} =  \frac{i}{2\pi\alpha'}\int d^2\sigma\sqrt{|\bar g|}
[\bar\theta^1P_-^{ij}\rho_i(D_j\theta)^1
+\bar\theta^2P_+^{ij}\rho_i(D_j\theta)^2]
%\equiv \frac{i}{2\pi\alpha'}\int d^2\sigma\sqrt{|\bar g|}
%\bar\theta^IA_F^{IJ}\theta^J,
\label{fermionic}
\end{equation}
where the Majorana condition is
\begin{equation}
\bar\theta^I = \tilde\theta^IC\times C.
\label{mw}
\end{equation}
Here $C$ is the $4\times 4$ charge conjugation matrix (see Appendix
B) and is a worldsheet  scalar.  The covariant derivative is given
by
\begin{equation}
D_j^{IJ}=\delta^{IJ}{\cal D}_j-\frac{i}{2L}\epsilon^{IJ}\rho_j,
\label{covd}
\end{equation}
where
\begin{equation}
{\cal D}_j=\frac{\partial}{\partial\sigma^j}+\frac{1}{4}\eta_j^\mu
(\Omega_\mu^{ab}\gamma_{ab}+\Omega_\mu^{a'b'}\gamma_{a'b'}),
\label{calD}
\end{equation}
with $\Omega_\mu^{ab}$ and $\Omega_\mu^{a'b'}$ the spin connections of the target spacetime.  Note that, to quadratic order in the fermions, this action
has the same functional form as that obtained in \cite{Metsaev:1998it} in
the vacuum AdS$_5\times S^5$ background.  Of course, the full Green-Schwarz
action in a non-trivial background is rather complicated, and it is not at
all clear whether the higher order in fermion terms would continue to match
up.  One way to address this question may be to return to the IIB superspace
formulation of the Green-Schwarz action in \cite{Grisaru:1985fv}.  However,
for the present work it is sufficient to note that the quadratic action
(\ref{eq:gsact}) with the supercovariant derivative given by (\ref{eq:scder})
is valid in an arbitrary 5-form background.

To proceed, we note that the fermionic action (\ref{fermionic}) is invariant
under the $\kappa$-transformation
\begin{equation}
\delta\theta^I=2i\bar\rho_j\kappa^{Ij},
\end{equation}
where $\kappa^{Ij}$ is a local spacetime spinor and a
worldsheet vector satisfying the duality relations
\begin{equation}
P_-^{ij}\kappa_j^1 = \kappa^{1i},\qquad
P_+^{ij}\kappa_j^2 = \kappa^{2i}.
\end{equation}
This can be verified with the aid of (\ref{clifford}) and (\ref{mw}) and the
properties of the projection tensors
\begin{equation}
P_\pm^{ij}=P_\mp^{ji},\qquad
P_\pm^{ik}P_\pm^{jl}=P_\pm^{il}P_\pm^{jk},\qquad
P_\pm^{ik}\bar g_{kl}P_\mp^{lj}=0.
\end{equation}
We now fix the $\kappa$-symmetry by choosing the gauge
$\theta^1=\theta^2\equiv\theta$.  The action (\ref{fermionic}) then reduces to
\begin{equation}
S_{\rm F}=\frac{i}{2\pi\alpha'}\int d^2\sigma(2\sqrt{|\bar g|}\bar g^{ij}
\bar\theta\rho_i{\cal D}_j\theta
-\frac{i}{L}\epsilon^{ij}\bar\theta\rho_i\rho_j\theta).
\label{fermionic2}
\end{equation}
We furthermore consider only string worldsheets that are embedded in
Schwarzschild AdS$_5$ (and which are localized on a point in $S^5$).
In this case, the worldsheet Dirac matrices reduce to
\begin{equation}
\bar\rho_i=\rho_i=\eta_i^\mu E_\mu^a\gamma_a.
\end{equation}

As is natural in the Green-Schwarz formulation, the action (\ref{fermionic2})
is for spacetime fermions $\theta$.  To reduce this to an action for
worldsheet fermions, we employ a local rotation to line up the first
two f\"unfbeins along the worldsheet directions
\begin{equation}
\eta_i^\mu E_\mu^\alpha=e_i^\alpha,\qquad
\eta_i^\mu E_\mu^p=0,
\label{prop12}
\end{equation}
with $\alpha=0,1$ and $p=2,3,4$. In this case
\begin{equation}
\bar g^{ij}\eta_i^\mu\eta_j^\nu=G^{\mu\nu}-E_p^\mu E^{p\nu}.
\end{equation}
In what follows, we shall use Greek letters $\alpha, \beta$ for the
first two Lorentz indexes, and Latin letters $p,q,r,s$ for the
remaining three.  Introducing the worldsheet  projection of the
target space spin connection,
$\Omega_j^{ab}\equiv\eta_j^\mu\Omega_\mu^{ab}$, it is
straightforward to verify that
\begin{equation}
\Omega_j^{\alpha\beta}=\omega_j^{\alpha\beta},
\label{prop3}
\end{equation}
where $\omega_j^{\alpha\beta}$ is the ordinary worldsheet  spin
connection. Furthermore, we have $\rho_j=e_j^\alpha\gamma_\alpha$,
and the Dirac operator may be rewritten as
\begin{equation}
\bar g^{ij}\rho_i{\cal D}_j=e^{\alpha j}\gamma_\alpha{\cal D}^\prime_j+D_1+D_2.
\label{general}
\end{equation}
Here
\begin{equation}
{\cal D}^\prime_j=\frac{\partial}{\partial\xi^j}
+\frac{1}{4}\omega_j^{\alpha\beta}\gamma_{\alpha\beta},
\end{equation}
and the additional terms $D_1$ and $D_2$ of (\ref{general}) are given by
\begin{equation}
D_1 = 2e^{\alpha j}\Omega_j^{\beta p}\gamma_\alpha\gamma_\beta\gamma_p,\qquad
D_2 = e^{\alpha j}\Omega_j^{pq}\gamma_\alpha\gamma_{pq}.
\end{equation}

In general, the $D_1$ and $D_2$ terms could be nontrivial.  However,
for the class of minimal surfaces that we are interested in, the spin
connections satisfy
\begin{equation}
\Omega_j^{pq}=0,
\label{simp1}
\end{equation}
as well as the symmetry property
\begin{equation}
e^{\alpha j}\Omega_j^{\beta p}=e^{\beta j}\Omega_j^{\alpha p}.
\label{simp2}
\end{equation}
It immediately follows that $D_2=0$ and that $D_1=2e_\alpha^j
\Omega_j^{\alpha p}\gamma_p$.  Working out the covariant derivative of
$\eta_i^\mu E_\mu^a$, and using the static gauge property (\ref{prop12})
as well as the background equation of motion (\ref{EOM}), we may then
show that $D_1=0$.

Under the simplification $D_1=D_2=0$, the fermionic action reduces to
\begin{equation}
S_{\rm F}=\frac{i}{2\pi\alpha'}\int d^2\sigma\sqrt{|\bar g|}
(2e^{\alpha j}\bar\theta\gamma_\alpha{\cal D}^\prime_j\theta
-\frac{i}{L}\epsilon^{\alpha\beta}\bar\theta\gamma_\alpha\gamma_\beta\theta).
\end{equation}
We now take the representation where $\gamma_0=-i\sigma_1\times I_8$ and
$\gamma_1=\sigma_2\times I_8$ (with $\sigma_1$ and $\sigma_2$ being the
ordinary Pauli matrices).  Decomposing the $16\times 1$ Majorana spinor
$\theta$ into eight $2\times 1$ real spinors $\psi_n$, we finally obtain
the worldsheet fermion action
\begin{equation}
S_{\rm F}=\frac{i}{\pi\alpha'}\sum_{n=1}^8\int d^2\sigma\sqrt{|\bar g|}
(e^{\alpha j}\bar\psi_n\tau_\alpha \hat\nabla_j\psi_n
-\frac{i}{L}\bar\psi_n \sigma_3 \psi_n),
\end{equation}
where $\tau_0=i\sigma_1$, $\tau_1=\sigma_2$ and the worldsheet
covariant derivative is
\begin{equation}
\hat\nabla_j=\frac{\partial}{\partial\sigma^j}+\frac{1}{4}\omega_j^{\alpha\beta}
\tau_{\alpha\beta}.
\end{equation}
The fermionic path integral then gives
\begin{equation}
Z_{\rm fm}=\int[d\theta]e^{iS_F}=({\rm const.})(\det{A_{\rm F}})^4
=({\rm const.})(\det A_{\rm F}^2)^2,
\label{zf}
\end{equation}
where $A_{\rm F}$ is the $2\times 2$ operator
\begin{equation}
A_{\rm F}=e^{\alpha j}\tau_\alpha \hat\nabla_j-\frac{i}{L}\sigma_3,
\label{mf}
\end{equation}
with
\begin{equation}
A_{\rm F}^2=-\hat\nabla^j\hat\nabla_j+\frac{1}{4}R^{(2)}+\frac{1}{L^2}.
\end{equation}

Combining the bosonic and fermionic contributions (\ref{eq:zbs}) and
(\ref{zf}), we now obtain the complete one-loop partition function
\begin{equation}
Z=Z_{\rm bs}Z_{\rm fm}=\frac{(\det A_{\rm F}^2)^2}{(\det A_{\rm tr})^{1/2}}.
\label{final}
\end{equation}
In the next two sections, we will examine this partition function for the
cases of the long string and parallel lines Wilson loops.  Before proceeding,
however, we note that for a general embedding, the conditions (\ref{simp1})
and (\ref{simp2}) may not be satisfied. An example where $\Omega_j^{pq}\neq 0$
is given by the light-like parallel lines configuration underlying the
AdS/CFT dual calculation of the jet-quenching parameter measured at RHIC.
In such cases, we would end up with a fermionic action of the form
\begin{equation}
S_{\rm F}=\frac{i}{\pi\alpha'}\sum_{n,n'}\int d^2\sigma
\bar\psi_n[(\bar e^{\alpha j}\tau_\alpha \hat\nabla_j-\frac{i}{L}\sigma_3)\delta_{nn'}
+M_{nn'}]\psi_{n'}.
\end{equation}
%

%%%%%%%%%%%%%%%%%%%%%%%%%%%%%%%%%%%%%%%%

\section{A long string}

In this section, we evaluate the first quantum correction to the long
string solution in a Schwarzschild-AdS$_5$ background.  To be concrete,
we take the spacetime metric to be given by
\begin{equation}
ds^2=\fft{1}{z^2}\left[-f\,dt^2+d\vec X\,^2+\fft{dz^2}f\right]+d\Omega_5^2,
\label{AdSSchw}
\end{equation}
where
\begin{equation}
f=1-\fft{z^4}{z_h^4}.
\label{eq:fz}
\end{equation}
(Note that we have set the AdS$_5$ radius to unity, i.e. $L=1$.)  This metric
corresponds to a planar horizon black hole in Poincar\'e coordinates.  This
is of course the appropriate gravity dual of $\mathcal N=4$ super-Yang-Mills
gauge theory on $\mathbb R^1\times\mathbb R^3$ or $S^1\times\mathbb R^3$, where $S^1$ of
the latter corresponds to the Euclidean time of period $1/({\rm temperature})=\pi z_h$.
The $z$ coordinate is such that $z=0$ corresponds to the boundary of AdS$_5$ and $z=z_h$ the horizon.

The Riemann tensor in the AdS$_5$ sector may be written as
\begin{equation}
R_{\mu\nu\rho\lambda}=-(G_{\mu\rho}G_{\nu\lambda}-G_{\mu\lambda}G_{\nu\rho})
+C_{\mu\nu\rho\lambda},
\end{equation}
where the nonzero components of the Weyl tensor are (up to symmetries)
\begin{eqnarray}
&&C_{0i0j} = \frac{f}{z_h^4}\delta_{ij},\qquad
C_{4i4j} = -\frac{1}{z_h^4f}\delta_{ij},\nonumber\\
&&C_{0404} = -\frac{3}{z_h^4},\kern2.3em
C_{ijkl} = \frac{1}{z_h^4}
(\delta_{ik}\delta_{jl}-\delta_{il}\delta_{jk}).
\label{eq:weylt}
\end{eqnarray}
Here $i,j,k,l$ labels the coordinates $X^1$, $X^2$ and $X^3$, while $X^0=t$
and $X^4=z$.

When computing both bosonic and fermionic fluctuations, we introduce the
natural f\"unfbeins of the spacetime metric (\ref{AdSSchw})
\begin{equation}
E^0=\fft1z\sqrt{f}dt,\qquad E^1=\fft1zdX^1,\qquad E^2=\fft1zdX^2,\qquad
E^3=\fft1zdX^3,\qquad E^4=\fft1{z\sqrt{f}}dz.
\label{5bein1}
\end{equation}
Since the worldsheet configurations that we are interested in reside at
a single point in $S^5$, we shall describe the fluctuations in $S^5$ in
terms of the cartesian coordinates of the $\mathbb R^5$ tangent space to
this point. These coordinates will be labeled by the letter $s$
with $s=5,6,7,8,9$, and the corresponding f\"unfbeins $E_\mu^s$ correspond to
those of a unit radius $S^5$ (corresponding to our setting the AdS$_5$ radius
to unity).  Corresponding to this choice of f\"unfbeins, the nontrivial
components of the spacetime spin connection in the AdS$_5$ sector are
\begin{equation}
\Omega_t^{04}=-\frac{1}{z}\Big(1+\frac{z^4}{z_h^4}\Big),\qquad
\Omega_j^{a4}=-\frac{\sqrt{f}}{z}\delta_j^a.
\label{eq:stOmega}
\end{equation}
with $f$ is given in (\ref{eq:fz}). The spin connection in the S$^5$ sector vanishes.

\subsection{The classical solution}

The long straight string corresponds to a worldsheet oriented along the
time and radial ($z$) directions.  It is a trivial solution of the
classical equation of motion (\ref{EOM}).  In static gauge, we take
\begin{equation}
\bar X^\mu(\tau,\sigma)=(\tau,0,0,0,\sigma;0,0,0,0,0).
\label{eq:lss}
\end{equation}
The induced worldsheet metric is given simply by the $t$ and $z$ components
of (\ref{AdSSchw})
\begin{equation}
ds^2=\frac{1}{\sigma^2}\Big(-fd\tau^2+\frac{1}{f}d\sigma^2\Big),
\label{induced1}
\end{equation}
where $f=1-\sigma^4/z_h^4$, and the corresponding worldsheet curvature
scalar is
\begin{equation}
R^{(2)}=-2\Big(1-\frac{3\sigma^4}{z_h^4}\Big).
\end{equation}

\subsection{Bosonic fluctuations}

To compute the bosonic fluctuations about the classical embedding
(\ref{eq:lss}), we introduce the fluctuation coordinates
$\xi^\mu(\tau,\sigma)\equiv\delta X^\mu(\tau,\sigma)$ along with their
tangent space counterparts $\zeta^{\hat a}=E_\mu^{\hat a}\xi^\mu$.
Here, $\zeta^0$ and $\zeta^4$ are longitudinal and may be set to zero in
static gauge.  Adding the transverse fluctuations to the background
configuration(\ref{eq:lss}), we have
\begin{eqnarray}
X^\mu(\tau,\sigma) &=& (\tau,\xi^1(\tau,\sigma),\xi^2(\tau,\sigma),\xi^3(\tau,\sigma),\sigma;
\xi^{s}(\tau,\sigma))\nonumber\\
&=&(\tau,\sigma\zeta^1(\tau,\sigma),\sigma\zeta^2(\tau,\sigma),
\sigma\zeta^3(\tau,\sigma),\sigma;\zeta^{s}(\tau,\sigma)).
\end{eqnarray}
This gives
\begin{equation}
dX^\mu=(d\tau,\dot\xi^1d\tau+\xi^1{}'d\sigma,\dot\xi^2d\tau+\xi^2{}'d\sigma,
\dot\xi^3d\tau+\xi^3{}'d\sigma,d\sigma;
\dot\xi^s d\tau+\xi^s{}'d\sigma),
\end{equation}
where $\dot{}=\partial/\partial\tau$ and $'=\partial/\partial\sigma$.
Substituting this into the Nambu-Goto action and expanding to the second order,
we obtain
\begin{equation}
S_{\rm NG}^{(2)}=\fft{1}{4\pi\alpha'}\int d^2\sigma\sqrt{|\bar g|}\left[
\sum_{a=1}^3\left(\bar g^{\tau\tau}(\dot\zeta^a)^2
+\bar g^{\sigma\sigma}(\zeta^a{}')^2+M^2(\zeta^a)^2\right)
+\sum_{s=5}^9\left(\bar g^{\tau\tau}(\dot\zeta^{s})^2
+\bar g^{\sigma\sigma}(\zeta^{s}{}')^2\right)\right],
\end{equation}
where we have assumed Dirichlet boundary conditions and have dropped the
surface terms associated with integration by parts. The metric coefficients
$\bar g^{\tau\tau}$ and $\bar g^{\sigma\sigma}$ refer to the line element
(\ref{induced1}).  The (position dependent) `mass' of the three fluctuation coordinates in the AdS$_5$ sector reads
\begin{equation}
M^2=2\Big(1+\frac{\sigma^4}{z_h^4}\Big)=\frac{8}{3}+\fft{R^{(2)}}3.
\end{equation}
while the fluctuations on $S^5$ have vanishing worldsheet masses.
It is now clear that the operator $A_{\rm tr}$ of (\ref{mtr}), which
governs the bosonic fluctuations, takes the form
\begin{equation}
A_{\rm tr}=\left(\matrix{(-\nabla^2+\fft83+\fft13R^{(2)})I_3 && 0\cr
0 && -\nabla^2I_5\cr}\right).
\label{bslong}
\end{equation}
We may obtain the vacuum AdS solution by taking the limit $z_h\to\infty$.
In this case, $R^{(2)}\to-2$, and this reproduces the result of
\cite{Drukker:2000ep} for the BPS configuration of the long straight
string in AdS.

\subsection{Fermionic fluctuations}

Turning to the fermionic sector, since the worldsheet is oriented along the
$0$ and $4$ directions, the pullback of the f\"unfbeins (\ref{5bein1}) to
the worldsheet according to (\ref{prop12}) gives the zweibeins
\begin{equation}
e^0=\frac{\sqrt{f}}{\sigma}d\tau,\qquad e^1=\frac{1}{\sigma\sqrt{f}}d\sigma.
\end{equation}
In addition, (\ref{prop3}) yields the worldsheet  spin connection
\begin{equation}
\omega_\tau^{01}=\Omega_t^{04} \qquad \omega_\sigma^{01}=0,
\label{spinconn1}
\end{equation}
where $\Omega_t^{04}$ is given by (\ref{eq:stOmega}) with $z$ replaced by
$\sigma$.  It follows from (\ref{calD}) that
\begin{eqnarray}
{\cal D}_\tau &=& \frac{\partial}{\partial\tau}
-\frac{1}{2\sigma}\left(1+\fft{\sigma^4}{z_h^4}\right)\gamma^0\gamma^4
=\hat \nabla_\tau\times I_8,\nonumber\\
{\cal D}_\sigma &=& \frac{\partial}{\partial\sigma}=\hat\nabla_\sigma\times I_8,
\end{eqnarray}
where $\hat\nabla_i$ is the worldsheet covariant derivative acting on
spinors.  In other words, the additional $D_1$ and $D_2$ terms of
(\ref{general}) are absent, and the fermionic operator is given simply
by $A_F$ of (\ref{mf}).

Finally, substituting the bosonic (\ref{bslong}) and fermionic (\ref{mf})
operators into (\ref{final}), we obtain the one-loop partition function
for the long straight string in a Schwarzschild-AdS$_5$ background
\begin{equation}
Z=\frac{{\det}^2(-\hat\nabla^2+1+\frac{1}{4}R^{(2)})}
{{\det}^{\frac{3}{2}}(-\nabla^2+\frac{8}{3}+\fft13R^{(2)})\,
{\det}^{\frac{5}{2}}(-\nabla^2)}
=\frac{{\det}^2(-\nabla_+^2+1+\frac{1}{4}R^{(2)})
{\det}^2(-\nabla_-^2+1+\frac{1}{4}R^{(2)})}
{{\det}^{\frac{3}{2}}(-\nabla^2+\frac{8}{3}+\fft13R^{(2)})\,
{\det}^{\frac{5}{2}}(-\nabla^2)},
\label{straight}
\end{equation}
where we have factorized the fermionic determinant into its two diagonal (chiral)
components, {\it i.e.}~$\hat\nabla^2={\rm diag}(\nabla_+^2,\nabla_-^2)$. We have
\begin{equation}
\nabla_\pm^2=\bar g^{\tau\tau}\left(\frac{\partial}{\partial\tau}
\pm\frac{1}{2}\omega_\tau^{01}\right)^2 + +\frac{1}{\sqrt{|\bar
g|}}\frac{\partial}{\partial\sigma} \left(\sqrt{|\bar g|}\bar
g^{\sigma\sigma}\frac{\partial}{\partial\sigma}\right)
%+\barg^{\sigma\sigma}\frac{\partial^2}{\partial\sigma^2}
\label{chiralcomp}
\end{equation}
with $\bar g^{ij}$ given by (\ref{induced1}) and $\omega_j^{\alpha\beta}$
by (\ref{spinconn1}).

%%%%%%%%%%%%%%%%%%%%%%%%%%%%%%%%%%%%%%%%

\section{The Parallel lines Wilson loop}

The long straight string considered in the previous section
corresponds to the minimal surface extending from a single static
quark into the bulk of AdS$_5$.  The standard AdS/CFT computation of
the quark-quark potential involves a `parallel lines' Wilson loop
where a string worldsheet  is stretched between two quarks on the
AdS boundary.  In this section, we begin with an overview of the
classical solution, and then turn to the first quantum corrections.

\subsection{The classical solution}

For the parallel lines Wilson loop, we choose the quarks to be separated
in the $X^1$ direction in the spacetime metric of (\ref{AdSSchw}).  The
string minimal surface then stretches in the time direction and forms a
curve in the $X^1$-$z$ plane.  We choose the gauge
\begin{equation}
\bar X^\mu(\tau,\sigma)=(\tau,\sigma,0,0,z(\sigma);0,0,0,0,0),
\label{eq:embed}
\end{equation}
where $z(\sigma)$, which specifies the string profile, may be determined
by the classical equation of motion (\ref{EOM}).  To calculate the induced
metric, we note that
\begin{equation}
d\bar X^\mu=(d\tau,d\sigma,0,0,z'd\sigma;0,0,0,0,0).
\end{equation}
Using (\ref{AdSSchw}) for the spacetime metric, we find
\begin{equation}
ds^2=\fft{1}{z^2}\left[-f\,d\tau^2+\left(1+\fft{z'^2}f\right)d\sigma^2\right].
\label{wsmetric}
\end{equation}
The Nambu-Goto action then takes the form
\begin{equation}
S_{\rm NG}=\fft{{\cal T}}{2\pi\alpha'}\int d\sigma\sqrt{f/z^4+z'^2/z^4},
\end{equation}
where ${\cal T}$ is the time period.

The equation of motion for $z(\sigma)$ following from the above action is
\begin{equation}
z''=\fft{f_z}2-\fft{2f}z-\fft2zz'^2+\fft{f_z}fz'^2,
\label{eq:zeom}
\end{equation}
where
\begin{equation}
f_z=\fft{\partial f}{\partial z}=-4\fft{z^3}{z_h^4}.
\end{equation}
Although this equation is rather non-trivial, it admits a first integral
\begin{equation}
z'^2+f=\fft{z_0^4}{f_0}\fft{f^2}{z^4}.
\label{eq:firsti}
\end{equation}
Here $z_0$ is the maximum value of $z$ reached by the minimal surface.  This
corresponds to the value of $z$ where the string makes its closest approach
to the horizon.  We have also defined $f_0=f(z_0)=1-z_0^4/z_h^4$.  This
allows us to write the on-shell induced metric as
\begin{equation}
ds^2=\fft{1}{z^2}\left[-f\,d\tau^2+\fft{fz_0^4}{f_0z^4}d\sigma^2\right],
\label{curvsclr2}
\end{equation}
where we ought to keep in mind that $z(\sigma)$ is a solution to the
first order equation (\ref{eq:firsti}) with the Dirichlet conditions
$z\left(\tau,-\frac{r}{2}\right)=z\left(\tau,\frac{r}{2}\right)=0$.
The curvature scalar for this induced metric reads
\begin{eqnarray}
R^{(2)}&=&\fft{4(f-2)+2z'^2(2-3f)}{(f+z'^2)}\nonumber\\
&=& 2\biggl[\frac{4(z_h^4-z_0^4)}{z_0^4f^2}-\frac{4(z_h^4-z_0^4)}{z_0^4f}
+\frac{z_0^4-3z_h^4}{z_h^4}+\frac{z_h^4}{z_0^4}f\biggr].
\label{eq:plr2}
\end{eqnarray}

\subsection{Bosonic fluctuations}

To compute the bosonic fluctuations about the classical solution, we introduce
the fluctuation coordinates
\begin{equation}
\xi^\mu(\tau,\sigma)=\delta X^\mu(\tau,\sigma)
\end{equation}
and their tangent space projections $\zeta^{\hat a}={\cal E}_\mu^{\hat a}
\xi^\mu$.  While the string is oriented along the $0$, $1$ and $4$ directions
in the tangent space defined by the f\"unfbeins (\ref{5bein1}), a
simple rotation in the $1$-$4$ plane \cite{Forste:1999qn}
\begin{eqnarray}
\left(\matrix{{\cal E}_\mu^1\cr {\cal E}_\mu^4\cr}\right)
&=&\left(\matrix{\cos\phi&&\sin\phi\cr-\sin\phi&&\cos\phi}\right)
\left(\matrix{E_\mu^1\cr E_\mu^4\cr}\right),\nonumber\\
{\cal E}_\mu^{\hat a}&=&E_\mu^{\hat a}\qquad\mbox{for}\quad\hat a\ne 1,4,
\label{rotation}
\end{eqnarray}
with $\tan\phi={z^\prime}/{\sqrt{f}}$, allows us to align the longitudinal
worldsheet directions with the $0$ and $1$ directions in the modified
f\"unfbeins basis defined by $\mathcal E_\mu^{\hat a}$.  In particular, it
follows that
\begin{equation}
\eta_i^\mu{\cal E}_\mu^{\hat a}=0,\qquad\hat a=2,3,\ldots,9,
\end{equation}
where
\begin{eqnarray}
\eta_0^\mu &=& (1,0,0,0,0,0,0,0,0,0),\nonumber\\
\eta_1^\mu &=& (0,1,0,0,z^\prime,0,0,0,0,0),
\end{eqnarray}
are the tangent vectors on the worldsheet . Therefore a longitudinal
fluctuation takes the form
\begin{equation}
\xi_{\rm l}^\mu = (z\sqrt{f}\zeta^0,z\zeta^1\cos\phi,0,0,z\sqrt{f}\zeta^1,0,0,0,0,0)
\label{paralong}
\end{equation}
and is a zero mode of the quadratic action of the fluctuations as is shown in the
appendix A. A transverse fluctuation reads
\begin{equation}
\xi_{\rm tr}=(0,-z\zeta^4\sin\phi,z\zeta^2,z\zeta^3,z\sqrt{f}\zeta^4\cos\phi,\xi^s)
\label{paratr}
\end{equation}
and represent a physical degree of freedom. Without losing
generality, we choose the static gauge by  taking  the superposition
of (\ref{paratr}) and (\ref{paralong}), $\xi^\mu=\xi_{\rm
tr}^\mu+\xi_{\rm l}^\mu$, such that $\xi^0=\xi^1=0$. This amounts to
setting $\zeta^0=0$ and $\zeta^1=\zeta^4\tan\phi$ and  the amount of
computation is reduced.

Including the fluctuations parameterized this way, the classical embedding
profile (\ref{eq:embed}) is modified to
\begin{equation}
X^\mu(\tau,\sigma)=(\tau,\sigma,\xi^2(\tau,\sigma),
\xi^3(\tau,\sigma),z(\sigma)+\xi^4(\tau,\sigma),\xi^s(\tau,\sigma))
\end{equation}
with
\begin{equation}
\xi^2=z\zeta^2, \qquad \xi^3=z\zeta^3, \qquad
\xi^4=z\sqrt{f+z^{\prime 2}}\zeta^4 \,\, \hbox{and}\,\,
\xi^s=\zeta^s \label{transformation}
\end{equation}
so that
\begin{equation}
dX^\mu=(d\tau,d\sigma,\dot\xi^2d\tau+\xi^2{}'d\sigma,
\dot\xi^3d\tau+\xi^3{}'d\sigma,
\dot\xi^4d\tau+(z'+\xi^4{}')d\sigma;\dot\xi^s d\tau+\xi^s{}'d\sigma).
\end{equation}
In order to evaluate the Nambu-Goto action to second order in the fluctuations,
we must keep in mind that the $X^4$ coordinate is now $X^4=z+\xi^4$.  This
means that the metric function $f(X^4)$ ought to be expanded as
\begin{equation}
f(X^4)=f(z+\xi^4)=f+\xi^4f_z+\ft12(\xi^4)^2f_{zz}+\cdots,
\end{equation}
where for simplicity $f$ always denotes the function $f(z)$ given by
(\ref{eq:fz}) unless $f(X^4)$ is indicated explicitly.  Expanding
$\sqrt{|g|}$ to quadratic order in the small fluctuations is tedious
but straightforward.  The result is
\begin{eqnarray}
\sqrt{-g}&=&\fft{1}{z^2}\sqrt{f+z'^2}\biggl[1
+\fft12\xi^4\left(\fft{f_z}{f+z'^2}-\fft4z\right)+\xi^4{}'\fft{z'}{f+z'^2}
\nonumber\\
&&\kern4em+\fft{f}{2(f+z'^2)}((\xi^a{}')^2+z^2(\xi^{s}{}')^2)
+\fft{f}{2(f+z'^2)^2}(\xi^4{}')^2\nonumber\\
&&\kern4em-\fft1{2f}((\dot\xi^a)^2+z^2(\dot\xi^{s})^2)
-\fft1{2f(f+z'^2)}(\dot\xi^4)^2\nonumber\\
&&\kern4em+(\xi^4)^2\left(\fft3{z^2}+\fft{f_{zz}/4-f_z/z}{f+z'^2}
-\fft{f_z^2}{8(f+z'^2)^2}\right)\nonumber\\
&&\kern4em+\xi^4\xi^4{}'\left(-\fft{2z'}{z(f+z'^2)}
-\fft{f_zz'}{2(f+z'^2)^2}\right)\biggr].
\end{eqnarray}
The $a$ index corresponds to the $2$ and $3$ directions, while the $s$
index denotes tangent-space indices on $S^5$.

After an integration by parts, the linear term in $\xi^4$ gives the
equation of motion (\ref{eq:zeom}).  The quadratic terms are the
ones that we are interested in. Upon substitution of
(\ref{transformation}) and integration by parts for the terms
containing $\xi^a\xi^a{}'$ or $\xi^4\xi^4{}'$, we end up with
\begin{eqnarray}
S_{\rm NG}^{(2)}&=&\fft{1}{4\pi\alpha'}\int d^2\sigma\sqrt{|\bar g|}\lbrace
\sum_{a=2}^3\Big[\bar g^{\tau\tau}(\dot\zeta^a)^2
+\bar g^{\sigma\sigma}(\zeta^a{}')^2
-\fft{z}{2}\left(f_z-\fft{4f}z+\fft{f_zz'^2}{f+z'^2}\right)(\zeta^a)^2\Big]\nonumber\\
&&\kern8em+\bar g^{\tau\tau}(\dot\zeta^4)^2+\bar g^{\sigma\sigma}(\zeta^4{}')^2\nonumber\\
&&\kern8em+\fft{z}{2}\fft{1}{f+z'^2}\left(ff_z-f_z^2z+\fft{4fz'^2}z-2f_zz'^2+ff_{zz}z\right)(\zeta^4)^2\nonumber\\
&&\kern8em+\sum_{s=5}^9\Big[\bar g^{\tau\tau}(\dot\zeta^s)^2
+\bar g^{\sigma\sigma}(\zeta^s{}')^2 \Big]\rbrace,
\end{eqnarray}
where we have freely integrated by parts because of the Dirichlet
boundary conditions. The metric coefficients $\bar g^{\tau\tau}$ and
$\bar g^{\sigma\sigma}$ here correspond to the worldsheet line
element (\ref{curvsclr2}).  Just as in the long string case, the
fluctuating modes on $S^5$ have vanishing worldsheet masses
associated with them.  The remaining three transverse fluctuations
in AdS$_5$ have (position dependent) `masses'
\begin{eqnarray}
M_2^2=M_3^2&=&-\fft{z}2\left(f_z-\fft{4f}z+\fft{f_zz'^2}{f+z'^2}\right),
\nonumber\\
M_4^2&=&\fft{z}2\fft1{f+z'^2}
\left(ff_z-f_z^2z+\fft{4fz'^2}z-2f_zz'^2+ff_{zz}z\right).
\end{eqnarray}
Substituting in the explicit form of $f$ from (\ref{eq:fz}) then gives
\begin{eqnarray}
M_2^2=M_3^2&=&\fft{2f-2z'^2(f-2)}{f+z'^2}=2-\fft{2z'^2(f-1)}{f+z'^2},
\nonumber\\
M_4^2&=&\fft{8(f-1)-2z'^2(f-2)}{f+z'^2}=4+R^{(2)}
+\fft{4z'^2(f-1)}{f+z'^2}.
\end{eqnarray}

In the above, we have been able to simplify the expression for $M_4^2$
by substituting in the worldsheet curvature scalar (\ref{eq:plr2}).  Even
with this substitution, however, these individual masses have an extra term
\begin{equation}
\delta\equiv-\fft{2z'^2(f-1)}{f+z'^2},
\end{equation}
which may be related to the components of the Weyl tensor (\ref{eq:weylt})
according to
\begin{equation}
\delta=\eta^{\mu j}\eta^{\nu j}{\cal E}^{2\rho}{\cal E}^{2\lambda}
C_{\rho\mu\lambda\nu}
=\eta^{\mu j}\eta^{\nu j}{\cal E}^{3\rho}{\cal E}^{3\lambda}
C_{\rho\mu\lambda\nu}.
\end{equation}
This indicates that $\delta$ only vanishes in the vacuum AdS$_5$ geometry
when $f=1$ and the Weyl tensor vanishes identically.  Nevertheless, we
note that the masses obey the sum rule
\begin{equation}
\sum_iM_i^2=8+R^{(2)}.
\label{masssum}
\end{equation}
which also holds for the long straight string configuration of the
previous section.  Finally, this mass spectrum allows us to write the
bosonic fluctuation operator $A_{\rm tr}$ as
\begin{equation}
A_{\rm tr}=\left(\matrix{-\nabla^2+4+R^{(2)}-2\delta && 0 && 0\cr
                  0 && (-\nabla^2+2+\delta)I_2 && 0\cr
                  0 && 0 && -\nabla^2I_5\cr}\right).
\label{eq:plb}
\end{equation}
This reduces to the result found in \cite{Drukker:2000ep} in the vacuum
AdS$_5$ ({\it i.e.}~$\delta\to0$) limit.

\subsection{Fermionic fluctuations}

In order to consider the fermionic fluctuations, we start with the
worldsheet metric (\ref{wsmetric}) and choose the natural zweibeins
\begin{eqnarray}
e_i^0 =\frac{\sqrt{f}}{z}d\tau,\qquad
e_i^1 = \frac{1}{z}\sqrt{1+\frac{z^{\prime 2}}{f}}d\sigma.
\end{eqnarray}
The corresponding worldsheet  spin connection reads
\begin{equation}
\omega_\tau^{01}=-\frac{2\sqrt{f_0}z}{z_0^2f}
\left(1+\frac{z^4}{z_h^4}\right)z^\prime,\qquad \omega_\sigma^{01}=0.
\label{spinconn2}
\end{equation}

Just as we have oriented the bosonic worldsheet using the tangent space
rotation (\ref{rotation}), we do the same for the Green-Schwarz fermions.
In this case, the spinor representation of the above rotation is implemented
by the $16\times 16$ matrix ${\cal S}=e^{-\frac{1}{2}\phi\gamma^1\gamma^4}$
such that
\begin{eqnarray}
{\cal S}\gamma^0{\cal S}^{-1}&=&\gamma^0,\nonumber\\
{\cal S}\gamma^1{\cal S}^{-1}&=&\gamma^1\cos\phi+\gamma^4\sin\phi.
\end{eqnarray}
The projection of the spacetime spin connection post rotation can be read
off from the transformation of the covariant derivative (\ref{calD}).
Starting with
\begin{eqnarray}
{\cal D}_\tau &=& \frac{\partial}{\partial\tau}-\frac{1}{2z}
\left(1+\frac{z^4}{z_h^4}\right)\gamma^0\gamma^4,\nonumber\\
{\cal D}_\sigma &=& \frac{\partial}{\partial\sigma}
-\frac{1}{2z}\sqrt{f}\gamma^1\gamma^4,
\end{eqnarray}
we find
\begin{eqnarray}
{\cal S}^{-1}{\cal D}_\tau{\cal S} &=& \frac{\partial}{\partial\tau}
-\frac{1}{2z}\left(1+\frac{z^4}{z_h^4}\right)
(\gamma^0\gamma^1\sin\phi+\gamma^0\gamma^4\cos\phi),\nonumber\\
{\cal S}^{-1}{\cal D}_\sigma{\cal S} &=& \frac{\partial}{\partial\sigma}
-\frac{1}{2}\left(\phi'+\frac{\sqrt{f}}{z}\right)\gamma^0\gamma^4.
\end{eqnarray}
Therefore the nontrivial components of the worldsheet  projection of
the spacetime spin connection with respect to the rotated
f\"unfbeins are
\begin{eqnarray}
\Omega_\tau^{01} &=& -\frac{1}{z}\left(1+\frac{z^4}{z_h^4}\right)\sin\phi
=\omega_\tau^{01},\nonumber\\
\Omega_\tau^{04} &=& -\frac{1}{z}\left(1+\frac{z^4}{z_h^4}\right)\cos\phi,\nonumber\\
\Omega_\sigma^{04} &=& \phi'+\frac{1}{z}\sqrt{f}.
\end{eqnarray}
Comparing this with the general expansion (\ref{general}), we find that
$D_2=0$ and that the quantity $e^{\alpha j}\Omega_j^{\beta4}$ is diagonal
with respect to the indexes $\alpha$ and $\beta$.  In this case, the $D_1$
term vanishes following the general arguments given previously.  This can
also be verified explicitly with the aid of the equation of motion
(\ref{eq:zeom}) and the expressions for $\sin\phi$ and $\cos\phi$.
The Green-Schwarz fermionic action is then reduced to a worldsheet fermionic
action for eight two-component Majorana spinors, and hence the general
expression (\ref{mf}) for the fermionic operator $A_{\rm F}$ remains
valid in this parallel lines Wilson loop case.

To summarize, by combining the bosonic (\ref{eq:plb}) and fermionic
(\ref{mf}) expressions, we have obtained the one-loop partition
function for the parallel lines configuration in the Schwarzschild-AdS$_5$
background
\begin{equation}
Z=\frac{{\rm det}^2(-\nabla_+^2+1+\frac{1}{4}R^{(2)})
{\rm det}^2(-\nabla_-^2+1+\frac{1}{4}R^{(2)})}
{{\det}^{\frac{1}{2}}(-\nabla^2+4+R^{(2)}-2\delta)\,
{\det}(-\nabla^2+2+\delta)\,{\det}^{\frac{5}{2}}(-\nabla^2)},
\label{parallel}
\end{equation}
where $\nabla_\pm$ is give by the same expression as (\ref{chiralcomp}) but
with $\bar g^{ij}$ referring to the metric (\ref{curvsclr2}) and $\omega_j^{\alpha\beta}$
to the spin connection (\ref{spinconn2}).
%

%%%%%%%%%%%%%%%%%%%%%%%%%%%%%%%%%%%%%%%%

\section{Discussion}

In this paper, we have explored some general properties of string
worldsheet fluctuations in a Schwarzschild-AdS$_5\times S^5$
background. For the physically interesting cases of the worldsheets
of a long string on the boundary and of two parallel lines on the
boundary, we have expressed the fluctuation partition functions in
terms of a set of determinants of a $1\times 1$ Laplacian with a
mass term. The same method applies to a circular Wilson loop which
we did not address in this paper. Several issues remain to be
addressed, however, before these determinants may be evaluated.

There are two types of divergences associated to the Wilson loop
analysis. One is the UV divergence of the string $\sigma$-model underlying
the determinants in the partition function and the other is caused
by placing the physical 3-brane at the AdS boundary.  (The latter is already
present at the classical level.)  To address the first one, we regularize the
determinants with a heat kernel expansion, while the second one is regularized
by pulling the 3-brane slightly off the AdS boundary.  Using the heat kernel
expansion, we write
\begin{equation}
\ln\det\mathcal O = -{\rm Tr}\int_\epsilon^\infty\frac{dt}{t}e^{-t\mathcal O},
\end{equation}
where $\mathcal O$ is the operator $A_{\rm tr}$ of (\ref{eq:zbs}) or the
square of the operator $A_F$ of (\ref{fermionic}) and $\epsilon$ is a cutoff
parameter with dimensions of length squared.
We have \cite{Drukker:2000ep,Alvarez:1982zi}
\begin{equation}
\ln\det\mathcal O = \frac{a}{\epsilon}\int_W d^2\sigma\sqrt{\bar g}
+\ln\epsilon\left(b\chi_E+\int_W d^2\sigma\sqrt{\bar g}M^2\right)
+\hbox{finite terms},
\label{UVdiv}
\end{equation}
for Dirichlet or Neumann boundary conditions, where $W$ denotes the
string worldsheet , $M$ corresponds to the mass term of $\mathcal O$
and $a$, $b$ and $c$ are numerical constants. The Euler character of
the worldsheet  is given by
\begin{equation}
\chi_E=\frac{1}{4\pi}\int_W d^2\sigma \sqrt{\bar
g}R^{(2)}+\frac{1}{2\pi}\int_{\partial W}dsK,
\label{euler}
\end{equation}
where $\partial W$ is the boundary of the string worldsheet, $s$ is
the proper length along $\partial W$ and $K$ is the geodesic
curvature of $\partial W$.

The cancellation of the UV divergence of
(\ref{UVdiv}) for vacuum AdS$_5\times S^5$ with a general
embedding was discussed in \cite{Drukker:2000ep}, and the results
can be readily carried over to the Schwarzschild-AdS$_5\times S^5$
case. The $1/\epsilon$ term cancels between the bosonic and fermionic
determinants in the partition function.
The logarithmically divergent term proportional to the Euler character
cancels in the critical dimension $D=10$ independent of the details
of the target space. Finally, the cancellation of the logarithmic divergence
associated to the mass term requires the target space to satisfy the
IIB Einstein equation of motion,
\begin{equation}
R_{\mu\nu}=\fft12\frac{1}{2\cdot4!}F_{\mu\rho_1\rho_2\rho_3\rho_4}
F_\nu{}^{\rho_1\rho_2\rho_3\rho_4},
\end{equation}
which is the case for Schwarzschild-AdS$_5\times S^5$. The only temperature
effect, the nonzero Weyl tensor, does not contribute the Ricci tensor. Take
the parallel lines Wilson loop as an example: the contribution of the
$\delta$ terms from each determinant in the denominator of (\ref{parallel})
to the $\ln\epsilon$ term cancels in the product following the sum rule
(\ref{masssum}).

Turning to the reduced partition functions (\ref{straight}) and
(\ref{parallel}), one has to keep in mind a subtlety pointed out in
\cite{Drukker:2000ep}. The logarithmic divergence of the fermionic
determinant proportional to the Euler character is one quarter of
that extracted from the fermion action (\ref{fermionic2}) before
transforming the ten-dimensional Majorana-Weyl spinor $\theta$ into
worldsheet Majorana spinors because of the singularity of the
transformation.  However, this will not have an impact on the ratio
of two determinants with identical worldsheet  topology. For
example, the finite temperature correction to the heavy quark
potential, $\Delta V$ is given by the ratio of the fluctuation
determinant with a black hole to that without a black hole, and can
be evaluated unambiguously within the framework of (\ref{parallel}).
In particular, we have
\begin{equation}
V=-T\ln Z=-T(\ln Z_{\rm bs}+\ln Z_{\rm fm})=T\sum_n\left[\ln Z_{\rm
bs}(\omega_n) +\ln Z_{\rm fm}(\omega_{n+\frac{1}{2}})\right],
\end{equation}
where we have factorized the partition function into components of
different Matsubara frequencies through the substitutions of
$\frac{\partial}{\partial\tau} \to-\omega_n\equiv -2in \pi T$ for
bosons and $\frac{\partial}{\partial\tau}
\to-\omega_{n+\frac{1}{2}}\equiv -2i\left(n+\frac{1}{2}\right)\pi T$
for fermions. Using the Poisson formula
\begin{equation}
\sum_nf(n)=\int_{-\infty}^\infty dxf(x)+2{\rm Re}
\sum_{m=0}^\infty\int_{-\infty}^\infty dxe^{-2im\pi x}f(x),
\end{equation}
we find that
\begin{eqnarray}
V& =& -\int_{-i\infty}^{i\infty}\frac{d\omega}{2i\pi} \left[\ln
Z_{\rm bs}(\omega) +\ln Z_{\rm fm}(\omega)\right]\nonumber \\
& &+2{\rm
Re}\sum_{m=1}^\infty\int_{-i\infty}^{i\infty}\frac{d\omega}{2i\pi}
e^{-\frac{m\omega}{T}}\left[\ln Z_{\rm bs}(\omega)+(-1)^m\ln Z_{\rm fm}(\omega)\right]\nonumber \\
&=& -\int_{-i\infty}^{i\infty}\frac{d\omega}{2i\pi} \left[\ln Z_{\rm
bs}(\omega)+\ln Z_{\rm fm}(\omega)\right]
+\frac{2}{\pi}\int_0^\infty d\omega \left[\frac{{\rm Im}\ln Z_{\rm
bs}(\omega)}{e^{\frac{\omega}{T}}-1} -\frac{{\rm Im}\ln Z_{\rm
fm}(\omega)}{e^{\frac{\omega}{T}}+1}\right],
\end{eqnarray}
where in the last step, we have deformed the integration contour of
the second term on the RHS to wrap around the positive real axis along
which the singularities of $\ln Z_{\rm bs}(\omega)$ and $\ln Z_{\rm
fm}(\omega)$ reside. Only the first integral carries the logarithmic
divergence, which is cancelled by subtracting the heavy quark
potential at zero temperature, $T=0$,
\begin{equation}
V_0 = -\int_{-i\infty}^{i\infty}\frac{d\omega}{2i\pi} \left[\ln
Z_{\rm bs}^{(0)}(\omega)+\ln Z_{\rm fm}^{(0)}(\omega)\right].
\end{equation}
The difference $V-V_0$ may suffer from the second type of divergence
mentioned at the beginning of this section, {\it i.e.}\ the
divergence when the 3-brane is pushed back to the AdS boundary. This
divergence should be cancelled by subtracting the contributions from
two straight strings, as in the classical limit. The details of the
cancellation remains to be examined carefully  before the correction
to the screening length can be extracted.

Despite the simpler partition function of a straight string, the evaluation of
the partition function is actually more challenging in this case. Consider the
ratio of the partition function, $Z/Z_0$ with $Z$ given by (\ref{straight}) and
$Z_0$ its counterpart in the absence of the black hole. Although both world
sheets of $Z$ and $Z_0$ share the same boundary on the AdS boundary, their
extensions to the AdS bulk are quite different; one terminates at the
Schwarzschild horizon and the other at the AdS horizon.  The boundary condition
at the Schwarzschild horizon has to be examined carefully to guarantee the
divergence free condition of the ratio. Once the ratio is made cutoff
independent, dimensional arguments suggest that $Z/Z_0=1$.
Since $Z_0=1$ \cite{Drukker:2000ep}, this would imply that $Z=1$ as well.
This is, however,
not at all obvious, since the spectral equations of the operators
$-\nabla^2+\frac{8}{3}+\fft13R^{(2)}$ and $-\hat\nabla^2+1+\frac{1}{4}R^{(2)}$
appear highly nontrivial in the presence of a black hole. We hope to be able
to report our progress in this direction in the near future.

In the case of more complicated embeddings, such as the worldsheet
of boosted parallel Wilson lines, the Schwarzschild-AdS$_5\times
S^5$ metric in the rest frame of the corresponding heavy quarks
acquires cross terms and the operators underlying the functional
determinant may no longer be reduced to $1\times 1$ structures. This
also covers the case of light-like parallel Wilson lines that gives
rise to the jet quenching parameter from AdS/CFT duality
\cite{Liu:2006ug}. Also, in the latter case, the string worldsheet
becomes space-like with respect to real time, making it difficult to
define the path integral because of the non-positivity of the
fluctuation action. Thus we expect the evaluation of the $\mathcal
O(\lambda^{-1/2})$ corrections to the jet quenching parameter and
the drag force arising from worldsheet fluctuations to be somewhat
of a challenge. However, even without a complete computation, it
would be of interest to obtain at least a qualitative estimate of
the size of such corrections and whether they result in an
enhancement or suppression of the drag force.

%%%%%%%%%%%%%%%%%%%%%%%%%%%%%%%%%%%%%%%%
\acknowledgments

We thank H. Liu, L. Pando Zayas and A. Tseytlin for valuable discussions.  DFH
acknowledges the hospitality of the Institute of Nuclear Physics
at the University of Washington where part of this work was
completed. The work of DFH and HCR is supported in part
by NSFC under grant Nos.~10575043 and 10735040. The work of
DFH is also supported in part by Educational Committee of China
under grant NCET-05-0675 and project No.~IRT0624.  JTL
acknowledges the hospitality of the Institute of Particle Physics
at Huazhong Normal University where part of this work was
completed. This work was supported in part by the US Department of
Energy under grant DE-FG02-95ER40899.

%%%%%%%%%%%%%%%%%%%%%%%%%%%%%%%%%%%%%%%%

\appendix

\section {}

In this appendix, we present the details of the proof that a longitudinal
fluctuation is a zero mode of the Nambu-Goto action. The mode equation
with an eigenvalue $E$ reads
\begin{equation}
K_{1\mu}+K_{2\mu}=E\delta X_\mu,
\end{equation}
where $K_1$ and $K_2$ arise from the variation of $I$ and $J$ of (\ref{I1I2}),
respectively.  Explicitly,
\begin{eqnarray}
K_{1\mu}&=&-g^{ij}\nabla_i\nabla_j\delta X_\mu
-g^{ij}R_{\rho\mu\lambda\nu}\eta_i^\rho\eta_j^\lambda\delta X^\nu,\nonumber\\
K_{2\mu}&=&\frac{1}{2}(g^{ik}g^{jl}+g^{il}g^{jk}-g^{ij}g^{kl})
\nabla_i(\eta_{\mu j}\delta_1 g_{kl}),
\end{eqnarray}
where we have suppressed the overlines on the induced metric.
For a longitudinal fluctuation
\begin{equation}
\delta X^\mu = \eta_k^\mu\epsilon^k
\end{equation}
which generates a worldsheet  diffeomorphism
\begin{equation}
\delta_1g_{ij}=\nabla_j\epsilon_i+\nabla_i\epsilon_j,
\end{equation}
we find
\begin{eqnarray}
K_{1\mu} &=& -g^{ij}\Big[G_{\mu\nu}(\eta_k^\nu\nabla_i\nabla_j\epsilon^k
+2\nabla_i\eta_k^\nu\nabla_j\epsilon^k
+\epsilon^k\nabla_i\nabla_j\eta_k^\nu)
+R_{\rho\mu\lambda\nu}\eta_i^\rho\eta_j^\lambda\eta_k^\nu\epsilon^k\Big]
\nonumber\\
&=& -g^{ij}G_{\mu\nu}(\eta_k^\nu\nabla_i\nabla_j\epsilon^k
+2\nabla_i\eta_k^\nu\nabla_j\epsilon^k
+\epsilon^k[\nabla_i,\nabla_k]\eta_j^\nu)
+R_{\rho\mu\lambda\nu}\eta_i^\rho\eta_j^\lambda\eta_k^\nu\epsilon^k.
\end{eqnarray}
Here we have used the symmetry
\begin{equation}
\nabla_i\eta_j^\mu=\nabla_j\eta_i^\mu
\end{equation}
and the equation of motion (\ref{EOM}). It follows from the commutation
relation
\begin{equation}
[\nabla_i,\nabla_j]\eta_k^\mu
={R^\mu}_{\lambda\nu\rho}\eta_i^\nu\eta_j^\rho\eta_k^\lambda
-{R^{(2)l}}_{kij}\eta_l^\mu,
\end{equation}
that
\begin{equation}
K_{1\mu}=-g^{ij}G_{\mu\nu}(\eta_k^\nu\nabla_i\nabla_j\epsilon^k
+2\nabla_i\eta_k^\nu\nabla_j\epsilon^k+\frac{1}{2}R^{(2)}
\eta_i^\nu\epsilon_j).
\label{J1}
\end{equation}
The term $K_{2\mu}$ is simplified similarly. We have
\begin{eqnarray}
K_{2\mu} &=& (\nabla^i\eta^j_\mu)(\nabla_j\epsilon_i+\nabla_i\epsilon_j)
+\eta^j_\mu\nabla^i(\nabla_j\epsilon_i+\nabla_i\epsilon_j)
-\eta^i_\mu\nabla_i\nabla_j\epsilon^j\nonumber\\
&=& 2\nabla_i\eta^j_\mu\nabla_j\epsilon^i
+\eta^j_\mu\nabla_i\nabla^i\epsilon_j
+\eta^i_\mu[\nabla_j,\nabla_i]\epsilon^j\nonumber\\
&=& 2\nabla_i\eta^j_\mu\nabla_j\epsilon^i
+\eta^j_\mu\nabla_i\nabla^i\epsilon_j
-{R^{(2)j}}_{kij}\eta^i_\mu\epsilon^j\nonumber\\
&=& 2\nabla_i\eta^j_\mu\nabla_j\epsilon^i
+\eta^j_\mu\nabla_i\nabla^i\epsilon_j
+\frac{1}{2}R^{(2)}\eta^i_\mu\epsilon_i.
\label{J2}
\end{eqnarray}
Finally, by adding (\ref{J1}) and (\ref{J2}) together, we end up with
\begin{equation}
K_{1\mu}+K_{2\mu}=0,
\end{equation}
and hence the longitudinal fluctuation is indeed a zero mode.

\section{}

Here we review our spinor conventions.  Following
\cite{Drukker:2000ep,Metsaev:1998it}, we use the letters $a$, $b$, $c$, $\ldots$
to label the AdS$_5$ dimensions and $a^\prime$, $b^\prime$, $c^\prime$,
$\ldots$ to label the $S^5$ dimensions. The $32\times 32$ Dirac matrices
in $D=10$ may be written as
\begin{eqnarray}
\Gamma^a &=& \gamma^a\times I_4\times\sigma_1,\nonumber\\
\Gamma^{a'} &=& I_4\times\gamma^{a^\prime}\times\sigma_2,
\end{eqnarray}
where the $4\times 4$ gamma matrices $\gamma^a$
and $\gamma^{a^\prime}$ satisfy the anticommutation relations
\begin{eqnarray}
\lbrace\gamma^a,\gamma^b\rbrace&=&2\eta^{ab}=(-,+,+,+,+),\nonumber\\
\lbrace\gamma^{a^\prime},\gamma^{b^\prime}\rbrace
&=&2\delta^{a^\prime b^\prime}=(+,+,+,+,+).
\label{anticomm12}
\end{eqnarray}
Taken together, this gives
\begin{equation}
\{\Gamma^{\hat a},\Gamma^{\hat b}\}=2\eta^{\hat a\hat b}=(-,+,\cdots,+),
\end{equation}
where the superscripts $\hat a$ and $\hat b$ label all ten dimensions. The
charge conjugation matrix reads
\begin{equation}
{\cal C}=C\times C\times i\sigma_2,
\end{equation}
where $C$, the four-dimensional charge conjugation matrix, satisfies the
conditions $\tilde C=-C$, $C\tilde\gamma^a C^{-1}=\gamma^a$ and
$C\tilde\gamma^{a'} C^{-1}=\gamma^{a'}$.

The ten-dimensional analog of $\gamma_5$ is
\begin{equation}
\Gamma^{11}=\Gamma^0\Gamma^1...\Gamma^9=I_4\times I_4\times\sigma_3.
\end{equation}
A Weyl spinor is an eigenstate of $\Gamma^{11}$:
\begin{equation}
\Gamma^{11}\Psi_\pm=\pm\Psi_\pm,
\end{equation}
and the Majorana condition reads
\begin{equation}
\bar\Psi=\tilde\Psi{\cal C},
\label{majorana}
\end{equation}
with $\bar\Psi=\Psi^\dagger\Gamma^0$. A positive chirality IIB Weyl spinor
takes the form
\begin{equation}
\Psi=\psi\times\left(\matrix{1\cr 0\cr}\right),
\end{equation}
where $\psi$ is a 16 component spinor. For two such spinors,
$\Psi$ and $\Psi^\prime$, we have
\begin{equation}
\bar\Psi^\prime\Gamma^a\Psi=\bar\psi^\prime\gamma^a\psi, \qquad
\bar\Psi^\prime\Gamma^{a^\prime}\Psi=i\bar\psi^\prime\gamma^{a^\prime}\psi,
\end{equation}
where $\gamma^a$ and $\gamma^{a^\prime}$ designate the $16\times 16$
matrices $\gamma^a\times I_4$ and $I_4\times \gamma^{a^\prime}$.
This 16-component formulation is adopted in this paper. In terms of a
16 component spinor $\psi$, the Majorana condition (\ref{majorana})
reads
\begin{equation}
\bar\psi=\tilde\psi C\times C,
\label{majorana16}
\end{equation}
where $\bar\psi=\psi^\dagger\gamma^0$.

%%%%%%%%%%%%%%%%%%%%%%%%%%%%%%%%%%%%%%%%

\end{document}